\documentclass[12pt]{article}
\begin{document}

\title{Remarks on the classical capacity of quantum channel \thanks{
Work partially supported by INTAS grant 00-738.}}
\author{A. S. Holevo}
\date{}
\maketitle

\begin{abstract}
A direct proof of the relation between the one-shot classical capacity and
the minimal output entropy for covariant quantum channels is suggested. The
structure of covariant channels is described in some detail. A simple proof
of a general inequality for entanglement-assisted classical capacity is
given.
\end{abstract}

\textbf{1.} Let $\Phi $ be a quantum channel in a Hilbert space ${\mathcal{H}
}$ of finite dimensionality $d$, $\bar{C}^{(1)}(\Phi )$ -- its one-shot
classical capacity defined as
\begin{equation}
\bar{C}^{(1)}(\Phi )=\max_{\pi _{j},S_{j}}\left[ H(\sum_{j}\pi _{j}\Phi
(S_{j}))-\sum_{j}\pi _{j}H(\Phi (S_{j}))\right] ,  \label{cap}
\end{equation}
where $\left\{ \pi _{j}\right\} $ is a probability distribution, $S_{j}$ are
states in ${\mathcal{H}}$, $H(\cdot )$ is the quantum entropy. The relation
\begin{equation}
\bar{C}^{(1)}(\Phi )=\log d-\min_{S}H(\Phi (S)),  \label{cae}
\end{equation}
known to hold for bistochastic ($\Phi \lbrack I]=I$) qubit ($d=2$) channels
\cite{kr} , \cite{ahw}, was recently extended to a class of bistochastic
channels with $d>2$ \cite{cor}. We wish to remark that there is
a simple direct proof of (\ref{cae}) for substantially broader class of
channels.

Let $G$ be a compact group (in case $G$ is finite, it is equipped with the
discrete topology) and $g\rightarrow V_{g};g\in G,$ be a continuous
(projective) unitary representation of $G$ in $\mathcal{H}$. The channel $
\Phi $ is called \textit{covariant} with respect to the representation if
\begin{equation}
\Phi \lbrack V_{g}SV_{g}^{\ast }]=V_{g}\Phi \lbrack S]V_{g}^{\ast }
\label{covariant}
\end{equation}
for all $g\in G$ and all $S$.

\textit{Let the channel }$\Phi $\textit{\ be covariant with respect to an
irreducible representation, then the channel is bistochastic and (\ref{cae})
holds.}

Since the inequality $\leq $ is obvious, it is sufficient to show
\begin{equation}
\bar{C}^{(1)}(\Phi )\geq \log d-\min_{S}H(\Phi (S)).  \label{leq}
\end{equation}
Since the channel $\Phi $ is covariant, then $V_{g}\Phi \lbrack
I]V_{g}^{\ast }=\Phi \lbrack I]$ for all $g$. Irreducibility implies that $
\Phi \lbrack I]$ is proportional to $I$, hence is equal to $I$ by
normalization. Now assume first that the group is finite. From the
orthogonality relation for irreducible representations, we have (see, e. g. 
Sec. IV.2 of \cite{ho1})
\[
|G|^{-1}\sum_{g\in G}V_{g}SV_{g}^{\ast }=d^{-1}I.
\]
Take the state $S_{0}$ minimizing the entropy in (\ref{leq}). Then taking in
(\ref{cap}) states $S_{g}=V_{g}S_{0}V_{g}^{\ast };g\in G,$ with equal
probabilities $\pi _{g}=|G|^{-1}$ gives the value in the right hand side of
(\ref{leq}). If the group is continuous then similar argument applies, but
the optimizing distribution will be continuous, namely, the uniform
distribution on $G$. One can use then a finite approximation and continuity
properties to prove (\ref{leq}). $\triangle$

For example, (\ref{cae}) holds for the channel
\begin{equation}
\Phi \lbrack S]=\frac{1}{d-1}\left[ I-S^{T}\right] ,  \label{cas}
\end{equation}
which was used to disprove the multiplicativity conjecture \cite{hw}, since
it is covariant with respect to the action of the orthogonal group in ${\
\mathcal{H}}$. The conlusion remains valid for channels satisfying yet
broader covariance condition
\begin{equation}
\Phi \lbrack V_{g}^{(1)}SV_{g}^{(1)\ast }]=V_{g}^{(2)}\Phi \lbrack
S]V_{g}^{(2)\ast },  \label{cov1}
\end{equation}
where $V_{g}^{(1)},V_{g}^{(2)}$ are both irreducible representations.

The structure of covariant (conditionally) completely positive maps was
extensively studied in the context of covariant dynamical semigroups, see e.
g. \cite{h1} , \cite{h2}. In particular, for arbitrary covariant channel
there is the Lindblad representation $\Phi \lbrack
S]=\sum_{k}L_{k}SL_{k}^{\ast },$ where $L_{k}$ are the components of a
tensor operator for the group $G$, i. e. satisfy the equations
\[
V_{g}L_{j}V_{g}^{\ast }=\sum_{k}d_{jk}(g)L_{k},
\]
where $g\rightarrow \left[ d_{jk}(g)\right] $ is a unitary matrix
representation of $G.$ Tensor operators have been studied in detail, see e.
g. \cite{br}.

A complete description can be given for the Weyl-covariant channels \cite{h2},
the discrete version of which happens to be the class considered in a less
explicit form in \cite{cor}. Let $H$ be finite Abelian group, its dual $
\hat{H}\simeq H,$ and let $\alpha \rightarrow U_{\alpha },\alpha \in H;\quad
\beta \rightarrow V_{\beta },\beta \in \hat{H}$, be a couple of unitary
representations in a Hilbert space $\mathcal{H}$, satisfying the Weyl
canonical commutation relations
\[
V_{\beta }U_{\alpha }=\exp i\langle \beta ,\alpha \rangle U_{\alpha
}V_{\beta },
\]
where $\langle \beta ,\alpha \rangle $ is the duality form. Defining $
W_{z}=U_{\alpha }V_{\beta }$ for $z=(\alpha ,\beta ),$ we have a projective
unitary representation $z\rightarrow W_{z}$ of the additive group $G=H\oplus
\hat{H}$, namely
\[
W_{z}W_{z^{\prime }}=\exp i\langle \beta ,\alpha ^{\prime }\rangle
W_{z+z^{\prime }}.
\]
It follows
\begin{equation}
W_{z}^{\ast }W_{z^{\prime }}W_{z}=\exp i\left( \langle \beta ^{\prime
},\alpha \rangle -\langle \beta ,\alpha ^{\prime }\rangle \right)
W_{z^{\prime }}.  \label{ccr}
\end{equation}

Taking $H$ as a direct sum of cyclic subgroups $
\mathbf{Z}_{d_{1}},\dots ,\mathbf{Z}_{d_{s}}$, an (irreducible) Weyl
representation of the group $G=H\oplus \hat{H}$ is obtained by taking tensor
product of the (irreducible) representations of the subgroups $\mathbf{Z}
_{d_{j}}\oplus \mathbf{Z}_{d_{j}}.$ An irreducible representation of $
\mathbf{Z}_{d}\oplus \mathbf{Z}_{d}$ has the dimensionality $d$ and is
obtained from the ``generalized Pauli operators'' $U_{\alpha}, V_{\beta}$
which are discrete version of the Weyl operators in Schr\"odinger
representation.

\textit{A channel }$\Phi $\textit{\ is Weyl-covariant, }
\begin{equation}
\Phi \lbrack W_{z}^{\ast }SW_{z}]=W_{z}^{\ast }\Phi \lbrack S]W_{z},~~~z\in
G,\quad   \label{2.2}
\end{equation}
\textit{with respect to an irreducible representation} $z\rightarrow W_{z}$
\textit{if and only if }
\begin{equation}
\Phi \lbrack W_{z}]=\phi (z)W_{z}\mathit{\ },\quad   \label{2.3}
\end{equation}
\textit{where }$\phi (z)$\textit{\ is a positive definite function on }$G$.

From (\ref{2.2}), $\Phi \lbrack W_{z^{\prime }}]$ satisfies the same
relation (\ref{ccr}) as $W_{z^{\prime }}$, hence $\Phi \lbrack W_{z^{\prime
}}]W_{z^{\prime }}^{\ast }$ commutes with $W_{z},z\in G.$ Since the
representation $z\rightarrow W_{z}$ is irreducible,
\[
\Phi \lbrack W_{z}]=\phi (z)W_{z},
\]
where $\phi (z)$ is a complex function. From complete positivity of $\Phi $
it follows that $\phi (z)$ is positive definite. Indeed for any finite
collections $\{z_{j}\}\subset G,\{c_{j}\}\subset \mbox{\bf C}$
\[
\sum_{j,k}\bar{c}_{j}c_{k}\phi (z_{j}-z_{k})=\sum_{j,k}\bar{c}
_{j}c_{k}(W_{z_{j}}\psi |\Phi \lbrack W_{z_{j}}W_{z_{k}}{}^{\ast
}]W_{z_{k}}\psi )\geq 0,
\]
where $\psi \in \mathcal{H}$ is arbitrary unit vector.

Conversely, let the operator $\Phi \lbrack W_{z}]$ be defined by (\ref{2.3}).
There exists a random variable $\zeta $ in $\hat{G}=\hat{H}\oplus H$,
having the characteristic function
\begin{equation}
\phi (z)=\sum_{(x,y)\in \hat{G}}\exp i(\langle x,\alpha \rangle +\langle
\beta ,y\rangle )p_{x,y},\quad   \label{2.4}
\end{equation}
where $\left\{ p_{x,y}\right\} $ is the distribution of $\zeta $. Then for
all operators $X\mathcal{\ }$in $\mathcal{H}$
\begin{equation}
\Phi \lbrack X]=\sum_{x,y=1}^{d}p_{x,y}W_{(-y,x)}{}^{\ast }XW_{(-y,x)}=
\mathsf{E}W_{J\zeta }{}^{\ast }XW_{J\zeta },\quad   \label{2.5}
\end{equation}
where $J(x,y)=(-y,x),$ since for $X=W_{z}$ this follows from (\ref{ccr}), (
\ref{2.4}), and then can be extended by irreducibility. One easily sees that
the right hand side defines Weyl-covariant channel. $\triangle $

The relation (\ref{2.5}) suggests a dilation of the channel $\Phi $ to the
unitary stochastic evolution $S\rightarrow W_{J\zeta }{}^{\ast }SW_{J\zeta }$.
Such channels that are mixtures of unitary evolutions were characterized
in \cite{gw} as channels for which restoring of quantum information is
possible via measurement on environment. It was also observed in \cite{gw}
that while every such channel is bistochastic, the converse is not true,
namely, (\ref{cas}) is not a mixture of unitary evolutions for $d>2$.

\textbf{2. }In \cite{h3} we stated the inequality
\[
C_{ea}\left( \Phi \right) \leq \log d+\bar{C}^{(1)}(\Phi )
\]
for the entanglement-assisted classical capacity (see \cite{bbst}, \cite{h3}
)
\begin{equation}
C_{ea}\left( \Phi \right) =\max_{S}\left[ H(S)+H(\Phi \lbrack S])-H(S,\Phi )
\right] .  \label{ceac}
\end{equation}
A proof given in \cite{fan} follows from the inequality for the entropy
exchange $H(S,\Phi ):$\textit{\ if }$S=\sum_{j}p_{j}S_{j},$\textit{\ where }$
\left\{ p_{j}\right\} $\textit{\ is a probability distribution and }$S_{j}$
\textit{\ are pure states, then}
\begin{equation}
H(S,\Phi )\geq \sum_{j}p_{j}H(\Phi \lbrack S_{j}]),  \label{eex}
\end{equation}
by noticing that one can always take $S$ of this form in (\ref{ceac}). Here
we give shorter proof of (\ref{eex}).

From the definition, $H(S,\Phi )=H(S_{E}^{\prime })=H(\Phi _{E}[S]),$ where $
S_{E}^{\prime }$ is the final state of the environment, $\Phi _{E}$ is the
channel from the system $Q$ to the environment $E.$ By concavity of quantum
entropy, $H(\Phi _{E}[S])\geq \sum_{j}p_{j}H(\Phi _{E}[S_{j}]).$ But $H(\Phi
_{E}[S_{j}])=H(\Phi \lbrack S_{j}])$ since the final state of the system $QE$
starting in the pure state $S_{j}$ is again pure. $\triangle$

\end{document}